\documentclass[referee]{mn2e}
\usepackage{graphics}
\title[Infrared bands in emission]{Infrared lines, bands and plateaus in emission: from molecules to grains}
\author[R. Papoular]{R. Papoular$^{1}$\thanks{E-mail:
papoular@wanadoo.fr}\\
$^{1}$Service d'Astrophysique and Service de Chimie Moleculaire,
CEA Saclay, 91191 Gif-s-Yvette, France}
\begin{document}

\date{Accepted . Received ; in original form }

\pagerange{\pageref{firstpage}--\pageref{lastpage}} \pubyear{2002}

   \maketitle
\label{firstpage}

\begin{abstract}
Different approaches to the infrared emission spectrum of molecules are discussed with a view to understanding the spectral profile of the Unidentified Infrared Bands. Normal mode analysis, which delivers the integrated band intensities, is found to be inadequate: in particular, it does not account for the band widths. A more accurate procedure requires the computation of the energy contents of the vibration modes as well as the variations of the total electric dipole moment of the excited molecule. These quantities reflect the anharmonicity of atomic motions in molecules above the ground state, as a consequence of which neighboring modes interact and exchange energy. This has far reaching effects: a) the initial distribution of energy among modes, just after excitation, is far from uniform; b) energy is redistributed during radiative relaxation, so the energy ultimately radiated by a given mode differs from the energy that was initially imparted to it; c) vibration lines are broadened and acquire wings; d) as the molecule grows in size, line wings overlap and form plateaus underlying regions of high line concentration; e) this is accompanied by an increase of the number of phonons, vibrations involving the whole molecular structure, whose accumulation ultimately contributes to the far infrared continuum; f) the emission spectrum differs considerably from the normal mode spectrum and depends on the excitation process.  

This is illustrated by computations on molecules of increasing size: 6, 101 and 493 atoms. A tentative UIB-like synthetic spectrum is obtained by summing up the spectra of 21 CHONS structures, computed in the same way. Also discussed is the relative contributions of the molecule to band and continuum emission, as a function of its size.

\end{abstract}




\section{Introduction}

The quest for a realistic model of the structure, composition and excitation of the carriers of UIBs (Unidentified Infrared Bands) has been going on for more than three decades now. An essential requirement is, for such a model, to correctly reproduce the spectral features of UIB 
emission, which, by now, are very well documented in a large variety of astronomical sites. Despite some variations from object 
to object, there are enough similarities that it has been possible to partition most of the observed spectra into a small number
 of classes (see, for instance, Peeters et al. 2002, 2004; van Diedenhoven et al. 2004), thus allowing systematic comparisons to be made with laboratory or theoretical spectra.

One universal characteristic of UIBs is that they are relatively wide, and continuous even at a resolution of 1500; moreover, ``plateaus" appear beneath the stronger bands, between 6 and 8 $\mu$m and between 11 and 13 $\mu$m (see, for instance, Hony et al. 2001, Kwok and Zhang 2011, etc.); also, an underlying continuum may be seen to increase steadily from 6 $\mu$m upwards (see Smith et al. 2007). The present paper focuses on these particular aspects of UIBs.

Most present day candidate carriers are big molecules of widely different sizes (see Tielens, 2008), the size being 
limited, of course, by experimental or computational constraints. While the experimental production and spectral analysis of single macromolecules have proven to be daunting, modern computing capabilities make it easier to use theory or chemical simulation. The most usual computational way of determining the IR (InfraRed) spectrum of a given structure is to perform a Normal Mode Analysis (see Wilson et al. 1955). Once a composition and a structure have been chosen for the model carrier, the first step of this procedure is to compute the ground state in vacuum at 0 K, i.e. the state of lowest potential energy. A classical mathematical transformation then delivers all the different modes of atomic vibration of infinitesimal amplitude, characteristic of the grain. In the usual approximation, such a vibration corresponds to the motion of a massive point about the apex of a parabolic potential well. For a grain made up of N chemically bound atoms, there are 3N-6 such modes, each characterized by its frequency and the directions and velocities of the excursion of each atom from its equilibrium position (the apex), along the 3 space coordinates. From these, and the atomic charge distribution, the dipole moment can be deduced for each mode. Further treatment yields the mode \emph {IR intensity} (or \emph{line intensity}, or \emph{integrated band intensity}),
 which is proportional to the \emph {absorbance} at the corresponding frequency, the quantity commonly measured in the laboratory.

 While this procedure is higly instructive, it suffers, for our present purposes, from at least three defects:

a) The spectral mode intensity by itself is not sufficient to predict \emph {emission intensity} at the corresponding frequency. One also
 needs to know the extent to which this mode is kinetically excited; this, in turn, depends on the excitation process and on the coupling
 between modes.

b) As a result of excitation, emission lines may be shifted from absorption frequencies. 

c) By definition, the normal mode line intensity is a $\delta$ function and is therefore devoid of any width. By contrast, emission lines are broadened by anharmonicity of intramolecular atomic motions.

As early as 1992, Brenner and Barker \cite{bre} warned that ``unambiguous identification of the IS (InterStellar) emitter cannot be established just on the basis of matching the emission frequencies to laboratory absorption spectra". 
Brenner and Barker themselves computed the emission of excited benzene and naphtalene based on a) the ergodic assumption (statistical distribution of energy), b) Einstein's spontaneous emission coefficients, as deduced from measured (relative) absorption coefficients at a few characteristic spectral frequencies of the two molecules (see also Allamandola et al. 1989). Intrinsic bandwidth and underlying emission continuum were assumed.

The emission bandwidth may be the result of a number of processes acting together (e.g. Cook and Saykally 1998, Papoular 1999, Pech et al. 2002). Some width components depend on the energy content of the mode, 
and, therefore, on the excitation energy and the grain size. An essential broadening cause is the continuous exchange of energy between 
 modes coupled by anharmonicity. In this respect, comparison with laboratory measurements may be misleading. For, in the laboratory, the detection and spectroscopy of molecular bands usually require much higher densities and temperatures of molecules than found in interstellar space (e.g. Nemes et al. 1994, Frum et al.  1991), in which case \emph{pressure} broadening may far exceed other broadening effects (see, e.g., Kuhn 1971, Peach 1981). Another common broadening effect in the laboratory is molecular rotation (e.g. Pirali et al. 2009), which must be quite weak for large molecules in thin and cold astronomical environments. As for anharmonicity, it is not easy to determine, as it varies from vibration mode to vibration mode and from molecule to molecule (see Brenner and Barker 1992). For want of better knowledge of the broadening mechanisms in action, an arbitrary broadening is often convolved with computed normal mode intensities (e.g. Boersma a2010, b2011). An \emph{ad hoc} frequency shift may also be necessary to fit models with observations (e.g. Verstraete et al. 2001). The present state of the art in this domain is clearly summarized by Bauschlicher et al. \cite{bau} in their Appendix.

Here, these issues are addressed by using computational chemistry codes to \emph{deduce the emission intensity from the variations of the dipole moment of the particle after it has been excited}. In this way, the most important broadening processes are included, as well as the particulars of the excitation and relaxation processes, and the particle size. This method may be applied to vibrational emission following any type of excitation mechanism. No \emph{a priori} assumption (such as ergodicity or thermal equilibrium) is made regarding the relaxation process. The present work seeks to show why and how the discrete vibrational spectra of molecules evolve into continuous band spectra, as observed. It takes advantage of the huge progress that occured in digital computation chemistry during the last decades, which makes it possible, today, to apply long known theoretical notions to relatively large molecules. In a way, this work is also a complement, or a twin, to a previous paper describing in detail the excitation by H atom impact, and the ensuing rate of vibrational emission, regardless of the spectral profile (chemiluminescence; Papoular 2011a). 

To begin with, Sec. 2 describes three approaches to the spectrum, in order of increasing accuracy, as applied to a very simple, planar, molecule, ethylene C$_{2}$H$_{4}$. This highlights the strengths and  deficiencies of each approach. Attention is drawn to the finding that internal thermal equilibrium (or ergodicity) does not usually apply after excitation of the molecules, contrary to a common simplifying assumption (see, for instance, Brenner and Barker 1992, Mulas et al. 2006). 

The third, and more accurate, procedure, based on time analysis of the dipole moment variations, is then applied, in Sec. 3, to larger molecules (101 and 493 atoms), in which lines are broadened enough to form bands in crowded spectral regions. These molecules contain a variety of atoms of the ``CHONS" family (C,H,O,N, and S), some of the most abundant in space. Section 4 carries the endeavor a step further, by applying the dipole moment procedure to several more molecules of similar composition and 27 to 112 atoms in size each, and adding their emission spectra. As a result, plateaus, and even a budding continuum, become apparent. In Sec. 5, the expected evolution of the relative contributions of bands and continuum to emission is discussed as the molecules grow into grains by clustering or sticking. The main outcomes of this research are summarized in the Conclusion.

\section{Three approaches to vibrational spectra}

The present work is based on the use of various algorithms of computational organic chemistry, as embodied in the Hyperchem software provided
by Hypercube, Inc., and described in detail in their publication HC50-00-03-00, and cursorily, for astrophysical purposes, in 
Papoular \cite{pap01}. Here we use the
 improved version Hyper 7.5. While almost all the computations used standard algorithms, some required writing special purpose subroutines, a 
capability also provided by the software. Semi-empirical methods of quantum chemistry (PM3 and AM1) were used for molecules up to about 100 atoms in size. 
Beyond that, the less accurate, but faster Molecular Mechanics (MM+) method was preferred. Anharmonicity is built in and tailored by comparison with laboratory experiments on hydrocarbon molecules.

\subsection{Normal Mode Analysis}
Some of the main points of the present argument will now be made through chemical ``experiments" on ethylene. This was selected because it is small and, hence, amenable to more accurate simulation; the modes are few, far apart and easily distinguishable in frequency and behavior; and because
 the internal dynamics of its atomic vibrations were studied in detail in Papoular \cite{pap06}. The first step, in all approaches, 
is the building and optimization of the structure, i.e. minimizing its potential energy 
by automatic standard procedure.  This is found to be -533.63 kcal.mol$^{-1}$(in PM3; 1 eV=23 kcal.mol$^{-1}$). A standard NMA
 then delivers the integrated IR band intensity, $I(\nu)$, which is linked to the corresponding absorbance in cm$^{-1}$, at frequency $\nu$ 
(cm$^{-1}$), by

\begin{equation}
\alpha(\nu)=10^{2}I(\nu)\frac{C}{\Delta\nu}\,\,,
\end{equation}

where $\Delta\nu$ is the band width (cm$^{-1}$) and $C$ is the molecular density (mol.l$^{-1}$). The computation of $I$ is independent of $\Delta\nu$,
 and $\nu$ is the frequency of an ideal vibration of infinitesimal amplitude about the equilibrium atomic geometry. For ethylene, which has 12 modes,
the intensity spectrum is shown in Fig. 1.

 \begin{figure}
\resizebox{\hsize}{!}{\includegraphics{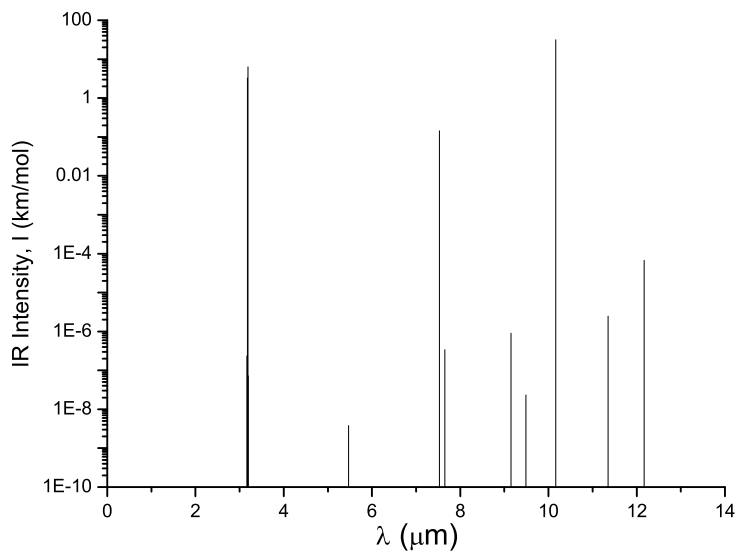}}
\caption[]{Infrared spectral intensities of ethylene C$_{2}$H$_{4}$. These are proportional to the corresponding absorption coefficients.
Note the logarithmic ordinate scale. In effect, only 3 modes have enough intensity for them to qualify as IR-active: 2 stretching modes, near 
3.2 $\mu$m, and the in-phase, out-of-plane CH bending mode at 10.2 $\mu$m. The others are ``dipole-moment forbidden".}
\end{figure}
             
By definition the NMA and spectral intensities are best suited to the prediction of the outcome of an absorption experiment, as this is done on 
samples which are (nearly) in their ground state. In the case of ethylene, only 3 of the modes are notably ``IR active": 2 (antisymmetric) CH 
stretching modes and 1 oop (out of plane), in phase CH vibration. Simple inspection of the molecular geometry shows that, in the other modes,
 the local dipole moments resulting from the movements of different atomic groups mostly cancel each other. This, of course, is a precious piece 
 of information. However, by itself, it is not sufficient to predict an emission spectrum. Obviously, this also depends on the energy content of 
each mode, which we take on next. 

\subsection{Energy distribution among modes}

An isolated molecule in space can aquire energy by, for instance, absorbing a photon, or by one of its dangling (C-) bonds capturing an H atom, thus increasing its potential energy by an amount equal to the binding energy (Papoular 2011a). In the following, we assume that the average interval between two successive excitation events is much longer than the IR lifetime of the molecule. So, the 
radiative \emph{power} of a molecule is the product of the frequency of exciting events and the \emph{energy} radiated by the molecule upon radiative relaxation. If the various vibration modes of the molecule were not coupled, all the energy deposited somehow in a particular mode would ultimately be reradiated at the frequency of the said mode. In fact, the modes are coupled by anharmonicity, so the modes which radiate efficiently, i.e. at a high rate, are constantly ``refueled" by those which radiate more slowly, until all the energy deposited by the exciting agent has been dissipated. Ultimately, therefore, the former modes will radiate more energy than the latter. Further analysis (see Appendix) shows that the total energy radiated by mode $\nu$ upon relaxation is proportional to

\begin{equation}
E^{*}(\nu)=\nu^{2}I(\nu)E_{0}(\nu),
\end{equation}

where $E_{0}(\nu)$ is the initial energy carried by the said mode, before radiative relaxation starts; this is assuming that the energy deposited by the exciting event is redistributed among the various modes at a much higher rate than the rate of radiative emission. Indeed this intramolecular   vibrational redistribution (IVR) is usually completed within a few nanoseconds, much shorter than the radiative lifetime (up to 1 s).

We therefore seek, now, to determine how the energy deposited in a carrier by the exciting agent is initially shared among the various modes. This partition,
$E(\nu)$, where the subscript 0 was dropped, has no direct link with the spectral intensities and may depend on the excitation process. Here, the focus is on excitation by H atom capture by a carbon dangling bond supposed to preexist at the surface of a hydrocarbon grain (Papoular 2011a).
 In this process, a free H atom approaches the grain closely enough to feel the attraction of a carbon atom whose bonds are incompletely occupied. 
Under favorable initial conditions, it falls into the corresponding potential well. It was shown that this capture reorganizes the distribution of 
the constitutive atoms in space so as to lower the ground state of
 the newly completed structure by an amount equal to the CH bond energy ($E_{b}\sim$4.5 eV). This is equivalent to depositing this energy in 
this new structure, an energy which is immediately and evenly divided into kinetic energy and potential energy (relative to the ground state of
the completed structure). The kinetic energy itself, initially localized in the CH stretching vibration, is then redistributed among
 this and other (not necessarily all) molecular vibrations. This could not occur were it not for the couplings between modes. While normal modes 
are uncoupled by definition, they become coupled as soon as some energy is deposited in the molecule, because of anharmonicity, which sets in 
above the ground state.

Note that, in this process, by contrast with UV photon excitation, the molecule remains in the \emph{electronic} ground state even though atomic vibrations are excited. To simulate this process on the computer, we start with an 
optimized structure (here, ethylene) \emph{at rest, with all bonds occupied}. One of the 4 CH bonds is then rotated out of the molecular plane 
by several degrees and extended from its original length by a few hundredths of \AA{\ } so as to increase the molecular potential energy by, say,  
$E$, which is tantamount to depositing this energy in the molecule. The molecule is then set free to perform a molecular dynamics run
 in vacuum, during which all atomic motions are monitored on the computer screen and recorded, as well as the total kinetic and potential 
energies, as a function of time. 

Initially, as expected, the dominant motion is the CH stretch accompanied by excursions out of the molecular plane. Soon, however, the other atoms
 are dragged along: inspection and comparison of the variations of, say, the bond lengths immediately reveals that energy
 constantly flows from bond to bond, thus increasing the amount of energy shared by other bonds. This process translates into a continuously 
decreasing standard deviation, $\sigma$, of each
 and every variable, like the total kinetic energy of the system, for instance. After some time (the relaxation time), a \emph{dynamic equilibrium} 
sets in,which is characterized by constant average relative standard deviations. The relative standard deviation of the energy, $\sigma_{E}/E$ is roughly $N_{M}^{-1/2}$, where $N_{M}$ is the number of effectively excited normal modes (see Reif 1984). The relaxation time is highly variable, depending on the grain structure and size, and on its initial perturbation. It may be as long as hundreds of picoseconds. This is still much shorter than the minimum radiative lifetime ($\sim1$ ms).

Once the dynamic equilibrium is reached, a new run is launched, during which all variables of interest are monitored. Here, we focus on the 
instantaneous total kinetic energy. Its FT (Fourier Transform) consists of peaks at twice the frequencies of those of the intensity spectrum,
 i.e. those of the normal modes (Sec. 2.1). What may vary (moderately) from run to run is the peak heights. These peaks are artificially
 broadened by the finite length, $\Delta\,t$, of the run: their width at half maximum (FWHM, cm$^{-1}$) is $\sim (c\Delta t)^{-1}$. The longer the run, the more representative is the spectrum, of the average energy content of the modes: 10 ps will do.

\begin{figure}
\resizebox{\hsize}{!}{\includegraphics{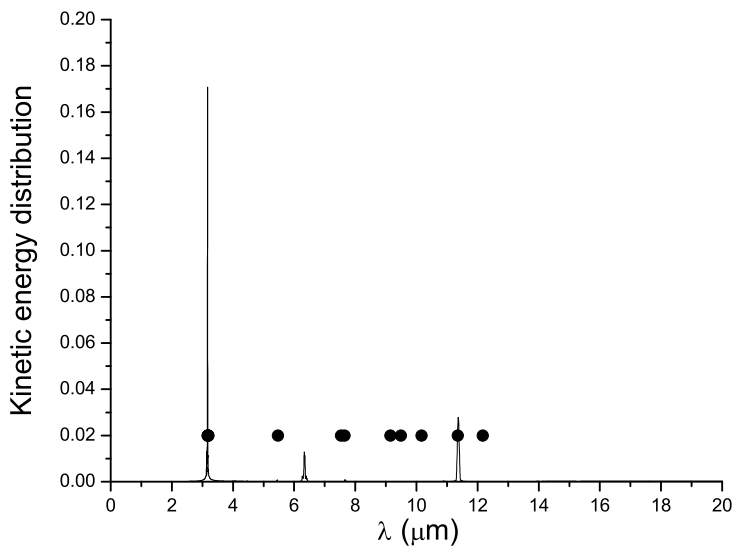}}
\caption[]{The curve illustrates the spectral distribution of molecular kinetic energy of ethylene after perturbation of one CH bond, represented
 here by the amplitude of the FT of the instantaneous kinetic energy variations during a 10 ps run of molecular dynamics on the free molecule
 in vacuum (see text). The dots mark the wavelengths of the normal modes. The average kinetic energy is 2.2 kcal/mol (vibrational ``temperature": 122 K). This spectrum is distinctly sparser than that of the IR spectral intensities, Fig. 1, because the normal modes are only weakly coupled. Nonetheless, this coupling gives rise to combination bands (intermodulation) between modes, which are visible upon zooming, but do not coincide with any of the dots.}
\end{figure}

Figure 2 illustrates the procedure as applied to the ethylene molecule, after it was energized as explained above: one of the CH bonds was rotated
 out of the molecular plane by 20 deg and extended from its original length, 1.086 \AA{\ }, to 1.15 \AA{\ } so as to increase the molecular potential energy
 by $E\sim5$ kcal/mol. After allowing some relaxation time, the total kinetic energy was recorded over a period of 10 ps. Its excursions from its average value, $\sim2.2$ kcal/mol, were extracted. The FT of the latter was delivered by a FFT (Fast Fourier Transform) algorithm, and its amplitude is
 represented in Fig. 2. This is a measure of the spectrum of the kinetic energy density; the scaling factor somewhat depends on the filter window used by the FFT algorithm, and on the length of the molecular dynamics run.

Only 2 significant peaks emerge in Fig. 2: one coinciding in wavelength with CH stretching normal modes, near 3.2 $\mu$m, and the o.o.p. (out of plane)  CH bending mode near 11.3 $\mu$m (cf. Fig. 1). The peak near 6.3 $\mu$m is a subharmonic of the strong stretching mode. To see if this result depends on the excitation process, a pristine molecule was first drowned in a thermal bath at 140 K so that it acquired an internal kinetic energy about equal to that injected by perturbing one of the CH bond (as in the previous experiment). It was then freed to vibrate in 
vacuum, as above. The resulting kinetic energy spectrum is shown in Fig. 3: although the stretching is still dominant, Fig. 3 is visibly different than Fig. 2, indicating that the spectrum does depend on the details of the excitation process; but there still is no energy 
equipartition between modes. Equipartition is approached only when the spectrum is taken on the molecule \emph{drowned within the bath, and continuously exchanging energy with it}, which, of 
course, is hardly the case when the molecule is in a thin medium, and excited only once in a while. This confirms that, at such low energy densities ($\sim0.2$ kcal/mode), in so small a molecule, the energy distribution in dynamic equilibrium, in vacuum, a short while after excitation, mainly depends on coupling between modes; and the vibration modes are only weakly coupled except for very close modes such as the CH stretchings. Indeed the records of the variations of the CH bond lengths, show continuous exchanges of energy between these bonds, not the others. 

\begin{figure}
\resizebox{\hsize}{!}{\includegraphics{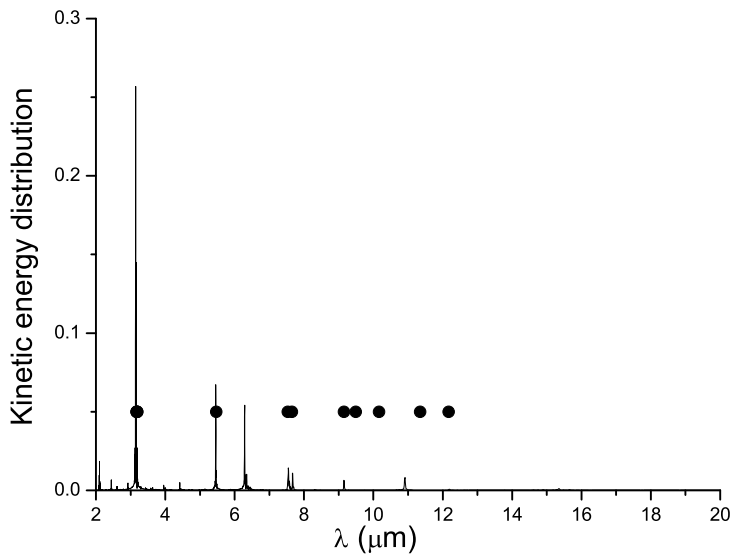}}
\caption[]{As in Fig. 2, the curve illustrates the spectral distribution of molecular kinetic energy of ethylene, represented by the amplitude of the FT of the instantaneous kinetic energy variations during a 10 ps run of molecular dynamics on the free molecule in vacuum (see text). Here, however, the excitation consisted in immersing the molecule in a thermal bath for a few picoseconds, until it reached \emph{thermal equilibrium} with about the same internal kinetic energy as in the previous experiment. It was then left free to vibrate in vacuum, and monitored as above, so as to obtain the energy spectrum. The dots mark the wavelengths of the normal modes. The spectrum is different, but not much richer, than that of Fig. 2.}
\end{figure}

These continuous exchanges translate into side bands on the wings of the 3.3 $\mu$ peak, which 
can be made visible by zooming on Fig. 2  and 3. For large molecules, these side bands contribute to forming underlying plateaus and continua in spectral regions of dense line clusters. A more detailed discussion of anharmonicity and mode coupling effects in this molecule can be found in Papoular \cite{pap06}. For our present purposes, note, 
in Fig. 2, the absence of the o.o.p., in phase, CH bending mode at 10.2 $\mu$m, which Fig. 1 shows to have the strongest IR intensity. This is 
because the asymmetry of the particular excitation process applied here is incompatible with the symmetry of this mode (all 4 CH bonds vibrating in phase), so the latter cannot be excited. As a consequence, if, using Fig. 1 and 2, a plot of $\nu^{2}I(\nu)E(\nu)$ is drawn to represent the spectral emission of the excited molecule (eq. 2),  it turns out that only two CH stretching modes emerge from the lot: the others either are too weakly excited or have weak IR intensity. However, as the $I'$s are $\delta$ functions at fixed wavelengths, this procedure runs into difficulties as soon as anharmonocity becomes strong enough to notably shift and/or broaden the spectrum of $E(\nu)$ w.r.t. that of $I(\nu)$. One is then led to the third, and more rigorous, procedure.

\subsection{Dipole moment analysis}

Chemical simulation experiments, such as described above, highlight the importance of anharmonicity as soon as energy is injected in a molecule. On the other hand, normal mode intensities are only valid in the ground state, or in the limit of infinitely small excursion from ground state. As a matter of fact, the $I'$s are specifically defined for \emph{absorption} from the ground state to the first excited state. Now, the electromagnetic \emph{emission} of a vibrating dipole is known to be proportional to the square of its electric dipole moment. A more
 realistic way of obtaining the emission spectrum in response to a given excitation is, therefore, to turn to the time variations of the electric dipole moment of our excited structure. Let this be $M(t)$, $<M(t)>$ its average and $\tilde{M}$ its variable part. Each frequency component, $m_{\nu}$, of $\tilde{M}$ corresponds to one normal mode of frequency $\nu$. As shown in the Appendix, it turns out that 

\begin{equation}
I(\nu)=2.6 \nu m_{\nu}^{2}\,\, ,
\end{equation}

where $I$ is in km/mol, $\nu$ in cm$^{-1}$ and $m$ in Debye.  It follows from eq. 2 that the radiated energy scales like 

\begin{equation}
E^{*}=\nu^{2}I(\nu)E(\nu)=3\,10^{-4}\nu^{3}m_{\nu}^{2}E(\nu)\,\, ,
\end{equation}

where $E(\nu)$, in eV, is the initial energy distribution, as obtained in Sec. 2.2.

   \begin{figure}
\resizebox{\hsize}{!}{\includegraphics{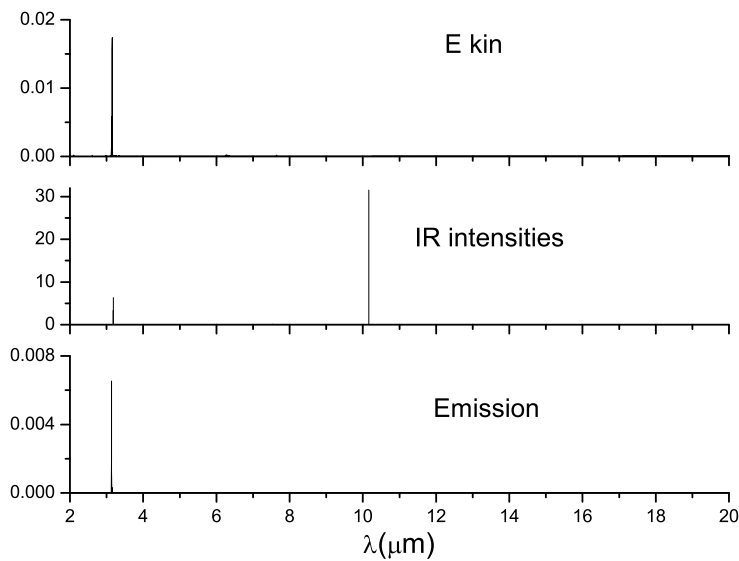}}
\caption[]{The bottom graph represents the spectrum of the energy or power radiated by the molecule after excitation by bond extension, as in Fig. 2. This was obtained by monitoring the total electric dipole moment of the molecule and applying eq. 4. For comparison, the middle graph represents the normal modes of Fig.1 (linear ordinates here), and the upper curve represents the kinetic energy spectrum of the excited molecule, as in Fig. 2. Most normal mode lines are conspicuously absent in emission , because the corresponding vibration modes are either IR-inactive or not excited.}
\end{figure}
   \begin{figure}
\resizebox{\hsize}{!}{\includegraphics{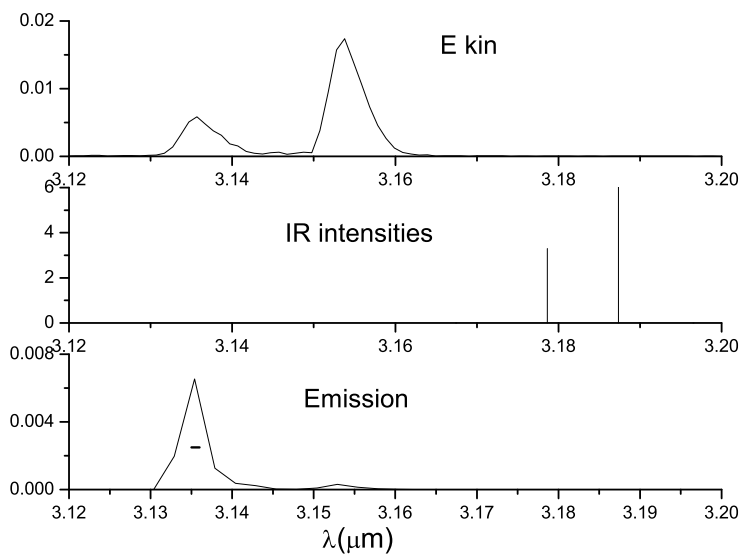}}
\caption[]{This is a zoom on the CH stretching region of Fig. 4. As a consequence of excitation and anharmonicity, the separation between the two IR-active CH-stretching lines is nearly twice that of the corresponding normal modes, and the shorter wavelength emission peak is dominant. The horizontal segment, of length 0.0012 $\mu$m (bottom graph), represents the artificial broadening due to the finite calculation run time (30 ps). The real, physical, broadening is $\sim$0.004 $\mu$m. The relative shift of $\sim0.035 \mu$m (1 $\%$) between the middle spectrum and the other two is probably to be ascribed to minor inconsistancies between the two distinct algorithms which compute normal modes and molecular dynamics, respectively.}
\end{figure}

The first expression of $E^{*}$ in eq. 4 is strictly valid for harmonic motions only, when $I(\nu)$ retains its original meaning, This is not the case of the second expression, for the dipole moment incorporates anharmonicity and broadening and shifting effects, if any. From now on, the latter expression will therefore be used to represent the radiation emission of the excited molecule. Now, chemical simulation codes also deliver $m(\nu,t)$: during a molecular
 dynamics run, these codes also monitor the displacement of electrical charges through the structure (using the Born-Oppenheimer approximation)
 and compute the 3 Cartesian components of the resulting overall dipole moment. To derive the squares of the individual frequency components
 of the latter, one
 computes the sum of the autocorrelation functions of the 3 components then takes the FT of this sum. This was done with the results of a 
run 10 ps in length (3333 cycles, 0.003 ps each in length; the dipole moment being computed after each cycle) and the bottom graph in Fig. 4 plots the product $\nu^{3}m_{\nu}^{2}E(\nu)$, to represent the radiation energy ultimately emitted (eq. 4), as a function of wavelength. For comparison, the middle graph represents the normal modes of Fig.1, and the upper curve represents the kinetic energy spectrum of the excited molecule, as in Fig. 2. Most normal mode lines are conspicuously absent in emission , because the corresponding vibration modes are either IR-inactive or not excited. 

Figure 5 is a zoom on the CH stretching region of Fig. 4. It reveals more subtle effects: as a consequence of excitation and anharmonicity, the separation between the two emission peaks of CH-stretching modes is twice that between the two corresponding normal mode peaks, and the ratio of the two emission peaks is reversed. Only dipole moment analysis can reveal such, and other, effects which go undetected by Normal Mode analysis or Kinetic Energy analysis alone.

\section{From lines to bands}

While the simple ethylene molecule is convenient for demonstrating the procedure and value of Dipole Moment analysis, it also hints at the fact that 
 small molecules (comprising less than a few tens of atoms) cannot carry the UIBs. This is also borne out by ample chemical simulation and observational evidence (see Kwok and Sandford 2008, Kwok and Zhang 2011). Obviously, this is because they cannot contain enough variety of sites and atomic species to match the variety of vibration features encountered in UIBs. 

In molecular spectra, it is possible to identify, \emph{in the near and mid-IR}, groups of vibration lines carried by, and characteristic of, small groups of bound atoms called \emph{functional groups}. Each of these is composed of the same chemical species associated in similar clusters, which may be replicated at different locations in the molecule. Depending on their environment, distinct functional groups of the same type will give rise to neighbouring but separate lines. One may therefore hope to build a \emph{band} by accumulating functional groups in \emph{slightly} different environments. Small molecules lack space to carry many distinct groups of the same type; so, even a very large number of identical small molecules will not do.

The situation improves with medium-sized (size of order 100 atoms) or large (of order 1000 atoms in size) molecules, as the lines crowd in each near- or mid-IR band. As the lines get closer, anharmonicity couplings become tighter. Those lines which are close enough to merge into bands are obviously carried by the modes of particular functional groups: CH stretchings, bendings (i.p. and o.o.p.), etc. The band forming process goes on with increasing molecular size, \emph{provided the average chemical composition and structure are preserved}. This is precisely how (disordered) solid state bands are formed. Besides, if the molecule is large enough, the wings of adjacent broadened lines encroach over each other, to form plateaus and, ultimately, an underlying continuum.

We therefore turn now to a more realistic model carrier, viz. a medium sized molecule, to monitor the spectral changes as the size increases. Figure 6 represents an organic ( so-called CHON) molecule comprising 51 carbon atoms, 39 hydrogen atoms, 8 oxygen atoms, 1 nitrogen and 2 sulfur atoms, for a total of 101 atoms. It is composed 
of 1 pyrene (4 pericondensed (compact) benzene rings) and 2 groups of 3 catacondensed (linear) rings including pentagonal rings. These 
substructures are linked together by CH$_{2}$ and O bridges. All three procedures described in Sec. 2 were applied. 

\begin{figure}
\resizebox{\hsize}{!}{\includegraphics{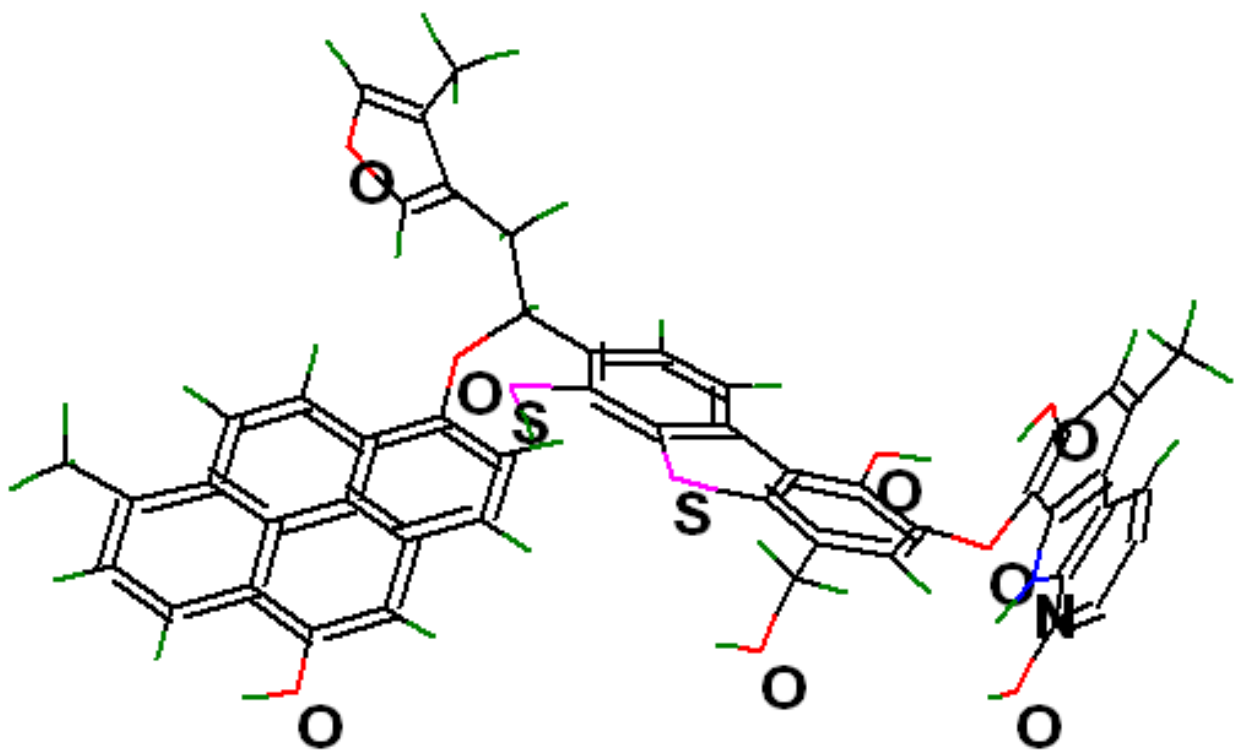}}
\caption[]{Aromatic and aliphatic structure with 101 atoms of the CHONS family. As usual, C and H atoms are not marked explicitely.}
\end{figure}

To simulate the capture of 
an H atom by a dangling (unoccupied) C bond as in chemilumunescence (see Introduction), the length of one of the CH bonds of the molecule was
 extended from 1.1 to 1.2 \AA{\ }. This 
increased the total potential energy by 104 kcal/mol (4.5 eV). The molecule was then allowed to relax freely in vacuum for 22 ps, during which
 potential energy progressively degenerated into kinetic energy until a dynamic equilibrium was reached, where both energies became nearly 
equal on average. The dipole moment spectrum was then obtained from a molecular dynamics run 10 ps in length, as in Sec. 2.

\begin{figure}
\resizebox{\hsize}{!}{\includegraphics{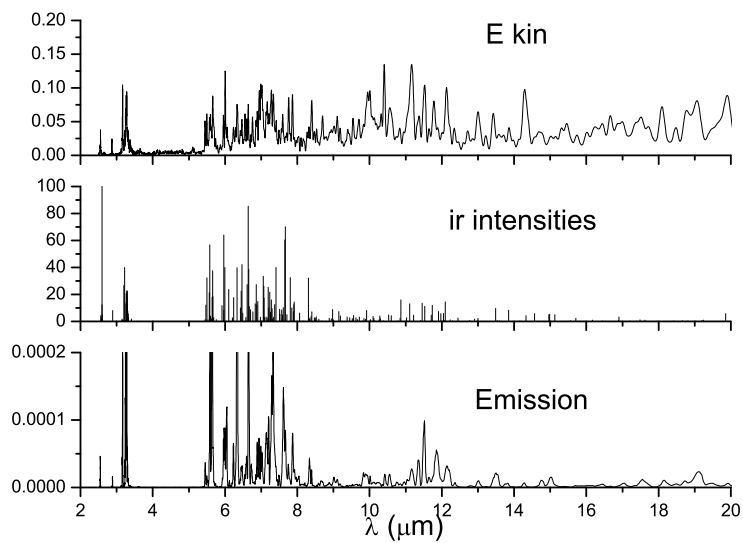}}
\caption[]{Each of the 3 graphs corresponds to one of the approaches (Sec. 2) to the energy radiated by the 101-atom molecule sketched in Fig. 6, when excited as described in the text. Each emission peak corresponds to a group of features appearing in \emph{both} the energy and the IR intensity graphs (cf. eq. 4): the emission spectrum is therefore sparser than both the other two. Note the budding continuum underlying the kinetic energy spectrum.}
\end{figure}

\begin{figure}
\resizebox{\hsize}{!}{\includegraphics{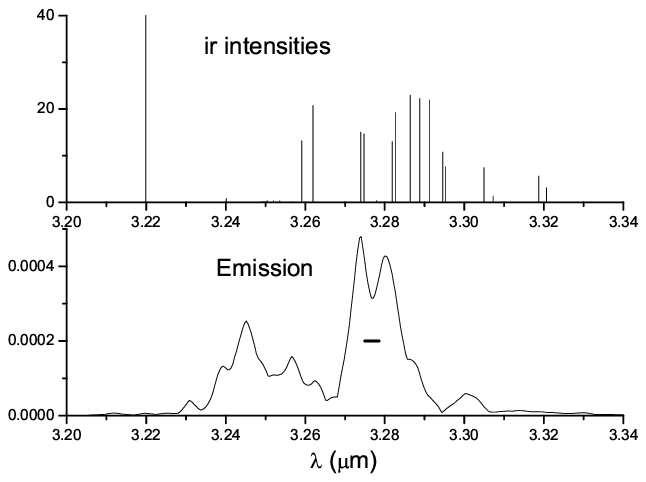}}
\caption[]{This is a zoom of Fig. 7, on a typical spectral region, with a resolution $\delta\lambda\sim0.0035\,\mu$m. The short horizontal segment represents this artificial broadening due to the finite calculation run time (10 ps). \emph{The much larger, physical, line broadening is not artificially enforced}. It is due to various effects of intrinsic anharmonicity revealed when the molecule is excited above the ground state. Also, close lines coalesce into bands. A related consequence is the piling up of wings of close lines, forming plateaus beneath the peaks  between 6 and 8$\mu$m, 11-12 $\mu$m, etc.}
\end{figure}

Figure 7 shows the normal mode spectrum (center graph, 297 lines) together with the energy distribution (upper graph) and the molecular dipole moment spectrum (bottom graph). It is observed again that many normal mode lines do not show up 
in the latter: they are either ``dipole forbidden" or simply not mechanically set in motion. Most importantly for our present argument, many
 \emph{lines} merge into \emph{bands}. This is illustrated in Fig. 8, by zooming in on the CH stretching region: the 58 CH stretching lines coalesce into essentially 2 bands, near 3.24 and 3.28 $\mu$m respectively, corresponding to CH bonds in different environments: O-capped pentagons and aromatics, respectively. Another example of this line coalescence is to be found in the CH o.o.p. bending region, 11-12.5 $\mu$m. Thus, bands were formed, whose width cannot be due to finite run time: the broadening due to the latter (10 ps here) is only 3.5$\,10^{-3}\,\mu$m. The cause is, rather, the fast and 
unceasing exchange of vibrational energy between close modes, which become coupled by anharmonicity as soon as energy is deposited in the molecule; 
this is amply documented by computational experiments (see for instance Papoular 2006). Here, this is made possible by the larger numbers of atoms, and hence of normal modes, so the latter come closer to each other.

The width of the aromatic CH stretching emission band at 3.28 $\mu$m is now 0.016 $\mu$m. This is not much narrower than Type 2 3.3-$\mu$m astronomical band of Tokunaga et al.\cite{tok} or van Diedenhoven \cite{vdied}, but still less than half his Type 1 (0.04 $\mu$m). A still larger molecule was therefore built by linking 5 copies of the 101-atom molecule of Fig. 6 through short bridges, to give a structure of 493 atoms. 
Using the PM3 semi-empirical method, Normal Mode Analysis delivered the spectrum of 1473 line intensities shown in Fig. 9, center graph.
 Such big molecular sizes are not easy to process with the semi-empirical algorithms designed for PC's, like Hyperchem: Fig. 9 required 720 h of calculation time. Performing Dipole Moment Analysis with the same chemical simulation method proved hopeless. It is, however, possible to resort
 to the less accurate, but much faster Molecular Methods, such as MM+ (see Papoular 2001), which is used hereafter. 

\begin{figure}
\resizebox{\hsize}{!}{\includegraphics{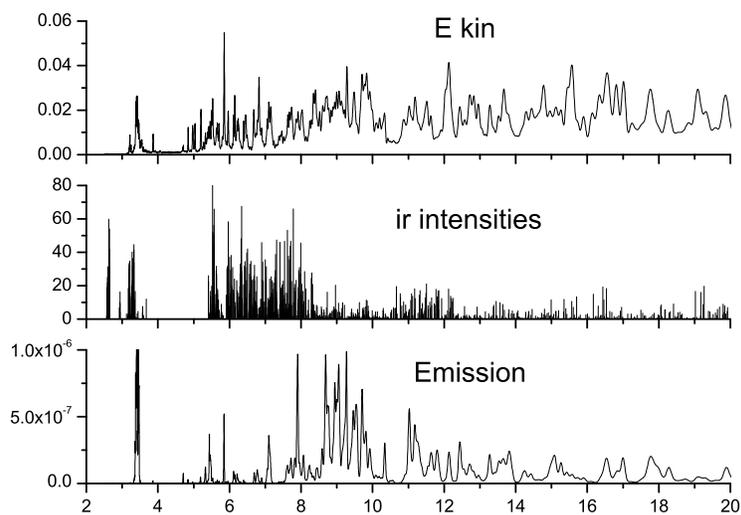}}
\caption[]{Each of the 3 graphs correspond to one of the approaches (Sec. 2) to the energy radiated by  a ``CHONS" molecule carrying 493 atoms, built up by linking together, through short bridges, 5 molecules, each identical with that shown in Fig. 6, and excited as described in the text. The same comments can be made as on Fig. 7.}
\end{figure}

\begin{figure}
\resizebox{\hsize}{!}{\includegraphics{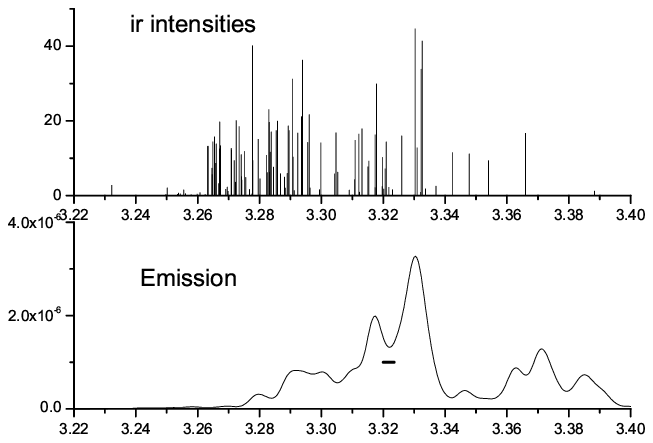}}
\caption[]{This is  a zoom of Fig. 9, on a typical spectral region, with a resolution $\delta\lambda\sim0.0035\,\mu$m, represented by the short horizontal segment. \emph{The much larger, physical, line broadening is not artificially enforced}. It is due to various effects of intrinsic anharmonicity revealed when the molecule is excited above the ground state. Also, close lines coalesce into bands. A related consequence is the piling up of wings of close lines, forming plateaus beneath the peaks  between 3.2 and 3.4 $\mu$m, 8.5 and 10 $\mu$m, 11 and 12 $\mu$m, etc.}
\end{figure}

This molecule was first excited, as above, by extending one of the CH bonds from 1.1 to 1.68 $\AA{\ }$, which gave it an excess potential energy equivalent to that which it would gain by capturing an H atom on a dangling bond (4.5 eV; see Sec. 2.2). After freeing it to relax in vacuum, a dynamic equilibrium was reached with a total internal kinetic energy of $\sim56$ kcal/mol, corresponding to a temperature of $\sim38$ K. A Dipole Moment Analysis was then performed on a molecular dynamics run of 3333 cycles, 0.003 ps each, for a total of 10 ps, ensuring negligible
 line broadening due to finite run time. The resulting spectra are shown in Fig. 9 and 10. As a result of the larger size of the molecule, the line broadening and coalescence signaled in Fig. 7 and 8 are only enhanced here. 

Another, related, consequence of the increased molecular size is the growth of \emph{plateaus} in regions of high normal mode density: between
 3.2 and 3.4 $\mu$m, 8.5 and 10 $\mu$m, 11 and 12 $\mu$m, etc. This is a result of the wings of close lines spreading out over each other.

 By comparison with Fig. 8, note the appearance, in Fig. 10, of new lines beyond 3.3 $\mu$m, due to new aliphatic sites in this molecule. This illustrates again how the number of close, but not coincident, IR lines can be made to increase in any spectral region characteristic of a functional group, so as to approach the shape of observed bands.

A potential problem with the Molecular Methods of chemical simulation, when it comes to absolute values, is the frequency error intrinsic to the code's use of empirical data optimized on molecules which necessarily differ from those of interest here. Thus, a difference of $\sim0.085\mu$m was noted between the IR intensity spectrum of Fig. 10 obtained with the semi-empirical method PM3 (center graph) and the emission spectrum (bottom graph) obtained with the MM code. In the latter figure, the spectrum was shifted in the opposite direction to restore agreement with the more accurate PM3 method.

As an aside, it must be stressed that the broadening demonstrated in Fig. 8 and 10, as well as the observed UIB widths, cannot be approached by any reasonable thermal Doppler broadening, as the velocity of heavy molecules in the cold astronomical environments is a tiny fraction of the velocity of light. Broadening by rotational bands must also be negligible, because of weight, cold and rarity of collisions.

\section{A tentative UIB-like synthetic spectrum}
Even allowing for line broadening, the line spectra of intermediate sized molecules (100-500 atoms) are not congested enough to yield continuous spectra resembling observed UIB spectra. One way forward is therefore to increase the molecular size. However, as stated above, a serious treatment of molecular emission becomes intractable beyond a size of 100 atoms or so. An alternative is to increase the number of tractable molecules and add their computed spectra. In this direction, an informed choice of adequate molecules can be guided by a study of the integrated absorption coefficients, $I(\nu)$, of a large number of CHONS molecules, which was done by Papoular \cite{pap11b}. The structures that were considered fall in general categories: aliphatic chains, aromatics, coronene-like, trios (a pentagon between 2 hexagons) and their associations. It was possible to associate each category with a characteristic congregation of absorption lines. Building upon this data, 21 structures were selected, ranging from 27 to 112 atoms in size, and their emission spectra calculated according to the procedure described above. Each was ascribed an adjustable weight, which was determined by cut-and-try so as to steer their weighted sum towards the shape of an arbitrary but characteristic UIB spectrum: that of NGC1482 (Smith et al. 2007). A preliminary result is drawn in Fig. 11.

\begin{figure}
\resizebox{\hsize}{!}{\includegraphics{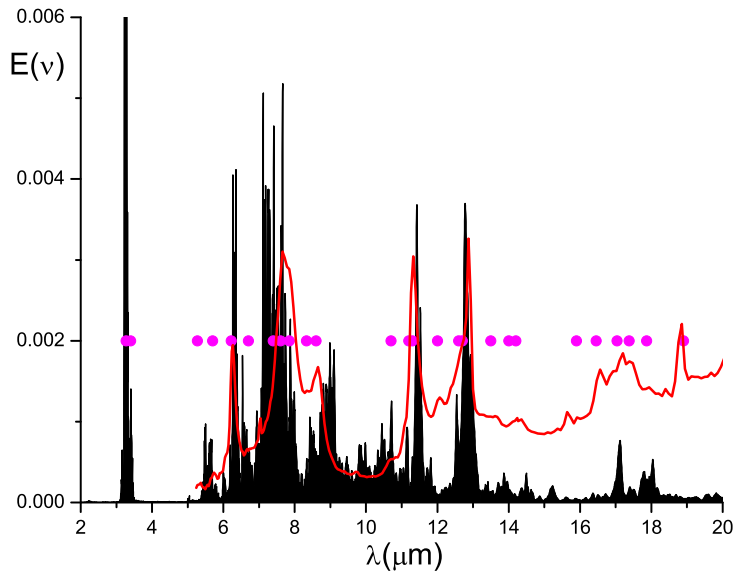}}
\caption[]{In black: the calculated emission spectrum, $E(\nu)$, of a collection of 22 CHONS structures described in the text and totalling 1110 atoms (3324 vibrational modes); the area under the curve is filled. Red line: for comparison, the measured flux from NGC1482 (Smith et al. 2007). Magenta dots: the detected features listed by Smith et al. (Table 3), plus the CH stretching features at 3.29 and 3.4 $\mu$m. All 3 sets of ordinates are arbitrary. Note the budding continuum between 6 and 14 $\mu$m, and the plateau between 16.5 and 18.5 $\mu$m. No line broadening is involved other than that which is due to anharmonicity and computed by the semi-empirical modeling code.}
\end{figure}

Even though no line broadening is involved here other than that due to finite time FFT (a negligible 0.0035 $\mu$m), and that which is due to anharmonicity and computed by the semi-empirical code, as explained above, the spectral intensity never falls to zero beyond 5 $\mu$m. Also, as expected, several \emph{bands} are formed where \emph{lines} congregate in the absorption spectra of the molecules. It must be stressed again that this result was obtained in the absence of any thermal, rotational or pressure broadening, or convolution of the lines with any arbitrary Gaussian or Lorentzian profile whose wings would produce similar effects.

 Improving the similarity of synthetic to observed UIB spectra requires the multiplication of molecules involved, as well as finer tailoring of their spectra by introducing small changes in their structures and in the details of their excitation. This is postponed to later work, as it falls outside the scope of the stated purpose of this paper.

\section{From molecules to grains}

At this stage, it has become clear that the synthesis of the various observed spectral emission bands and plateaus requires a very large number of medium-sized molecules or a smaller number of larger molecules, such as may provide the necessary diversity of functional groups, accompanied by  dense enough spectral line densities for the lines to start merging together under reciprocal attraction due to anharmonicity. On the other hand, if the observed underlying continuum is due to structures of the same composition as the said molecules, then these structures must be still larger. That large enough molecules exist is made possible by collisions in the ISM. Depending on the C/O ratio of the environment, silicate or carbonaceous grains will form preferentially. Both are subject to the mechanisms described above for the formation of bands (see Fig. 7 to 10). This evolution can be viewed as taking place in two steps. First, upon chance encounters, molecules can stick together through van der Waals attraction and H bonds, thus forming weakly coupled \emph{clusters}. As these grow larger with time, they become more resilient and covalent bonds may ultimately be formed within, giving finally birth to \emph{grains}.

How do the relative contributions of bands and continuum to dust emission vary as the grain size increases? Unfortunately, present common modeling packages can hardly handle clusters with thousands of atoms. It is possible, however, to forsee some of their general properties. Using the capabilities of available chemical simulation and visualization codes, one observes graphically that normal modes associated with functional groups flock below $\sim20\,\mu$m while modes above are associated with motions of a large part, or the whole, of the structure. The latter are called \emph{bulk} or \emph{skeletal} modes and often show up as ``slow" deformations of the structure, which can be identified with the \emph{phonons} of the solid state (so-called ``drumhead" modes). As the molecular size increases, phonons of proportionally longer wavelength are created, invading the far IR spectrum where they form a continuum.

According to Debye (see Rosenberg 1988), the upper limit of the phonon frequency in crystals is 

$\omega_{ph}=(6\pi^{2}N)^{1/3}v\,\,,$

where $N$ is the atomic density of the solid, and $v$, the sound velocity. Typically, on Earth, $N=10^{24}$ at.cm$^{-3}$, and $v=2000$ m.s$^{-1}$, leading to a Debye wavelength of $\sim20\,\mu$m, defining a lower limit to the photon spectrum. Moreover, Debye's elementary theory also shows that the spectral density of phonons increases as the square of their frequency, i.e. maximum density occurs at Debye's wavelength. This is at the origin of the frequent assumption that the far IR emissivity of grains goes like $\lambda^{-2}$ (see, for instance, Zhang et al. 2011).  

It must be kept in mind, however, that the UIB carriers cannot be as densely packed as earthly crystals are, for this would impede the formation of functional groups. So, Debye's limit in IS grains likely lies somewhat beyond 20 $\mu$m, as suggested by the spectral energy distributions emanating from circumstellar shells. But, in spite of a lower density, the relative contribution of grains to the near and mid-IR is bound to decrease as their size increases, for the phonons claim a growing fraction of the excitation energy: for the 493-atom molecule, this fraction is already 30 $\%$.

At the lower end of the size distribution, the contribution to UIB emission is limited by the condition that the structure of the functional groups not be perturbed severely enough by one exciting event, as to change the spectrum of their normal mode intensities beyond recognition. Let us arbitrarily define this limit by a maximum of two quanta per mode upon excitation. Assuming only half of the modes are effectively excited (see discussion in Sec. 2), and the weakest quantum energy to be 0.1 eV ($\lambda\sim13\, \mu$m), the limit is defined by

$E=0.2\,\frac{3N-6}{2}$,

where $E$ is the excitation energy in eV. For excitation by H capture, $E=4.5$ eV; for UV excitation, $E=20$ eV. The corresponding size ``cut-offs" are 17 and 69 atoms, respectively. Of course, these are to be considered only as orders of magnitude.

There is also the case of grains in thermal equilibrium with a luminous star in the vicinity. For such grains to be detected in the mid-IR, their steady-state temperature must be about 300 K. In this case, the average energy per atom, $E(eV)/N=T(K)/7800$, is too small to be of concern, whatever the grain size.

\section{Bands in absorption} 
One may wonder how our medium and large molecules behave in absorption. Unless they are maintained somehow in a state of permanent excitation, the molecules will remain in the ground state most of the time. It was stated in Sec. 2.1 that molecules in the ground state carried only discrete lines because the anharmonicity mechanisms are not active. Such lines can only be detected in absorption if their carriers are hugely concentrated, which is not the case in CS shells, and still less in the ISM. Apparently, no such line has yet been observed. The core of the difficulty probably lies in the stellar background continuum against which the absorption line must be observed. The fluctuations, $\Delta B$, of this background, $B$, set a lower limit to the detectable absorption signal. Indeed, let $\delta \nu$ be the spectral resolution of the detector, and $\Delta \nu$ the \emph{equivalent width} of the line to be detected. By definition of the latter,

$\Delta B \delta\nu= B \Delta\nu$.

But $\Delta\nu$ can also be expressed, in cm$^{-1}$, by

$\Delta\nu=\int\bf N\rm\sigma d\nu=\frac{\bf N\rm I}{6\,10^{18}}\,,$

where $\sigma$ is the absorption cross-section in cm$^{2}$, $I$ the IR intensity in km/mol, and $\bf N$, the line density of absorbers along the line of sight, in cm$^{-2}$. From this, the minimum line density of absorbers is deduced as
\begin{equation}
\bf N\rm=6\,10^{18}\frac{\delta B}{B}\frac{\delta\nu}{I}.
\end{equation}

For instance, if $\frac{\delta B}{B}=0.01\,,\delta\nu=10\,$cm$^{-1}$ and $I=100$ km/mol, then $\bf N\rm=6\, 10^{15}\,$cm$^{-2}$, which is hardly plausible in a CS envelope. 

For carriers to be observable in absorption, they have to be much larger still than those studied above, definitely the size and density of solid grains. For such objects, the functional groups that carry the UIBs will be destroyed or their features will be drowned in the continuum. This seems to be the case at least for the CH \emph{aromatic} stretching band which, to my knowledge, was never observed in absorption; instead, one observes a band extending from $\sim3.3$ to $\sim3.55\,\mu$m, characteristic of aliphatic CH stretchings in methine (CH), methylene (CH$_{2}$) and methyl (CH$_{3}$), both symmetric and asymmetric (see, for instance, Pendleton et al. 1994, Brooke et al. 1999). Figure 10 above suggests a tendency towards this direction of change. This is favored by the formation of dense internal webs of essentially aliphatic carbon structures. 

\section{Summary and conclusions}
In the laboratory, the IR absorption spectrum is the easiest route to understanding the structure of molecules; hence the universal use of \emph{IR intensity spectra}, $I(\nu)$, each \emph{line} of which corresponds to one of the normal modes of vibration. In space, detecting absorption lines requires unusually large concentrations of material; emission lines are much more common. It was shown above that absorption spectra are poor predictors of emission spectra. Not unexpectedly, they must be supplemented by the knowledge of the \emph{energy content of each vibration  mode} of the emitter. 

If the various normal modes remained isolated and strictly harmonic even in the excited state, then, each one would emit the whole energy that was deposited in it in the first place, regardless of its IR intensity. However, as soon as the molecule is energized and no longer in its ground state (in which it is, in the laboratory), \emph{anharmonicity} sets in, and induces internal \emph{couplings between modes}, which help redistribute the deposited energy. The equilibrium energy distribution among modes depends on the particular molecular excitation process (thermal, single UV photon absorption, capture of an H radical, etc.), but mostly on the internal couplings. Thus, for each candidate carrier of an astronomical spectrum, \emph{one cannot dispense with the computation of the dynamical equilibrium reached shortly after the deposition of excitation energy}. The analysis is made easier by the fact that the intramolecular energy redistribution is much faster than the IR emission process. 

During the emission process, the most IR active vibrations are energy-depleted faster than the others, which, in turn, refuel the former faster than they themselves radiate their own energy. In the end, it turns out (see Appendix), that a better approximation to the band emission scales with the product of initial energy content times the line intensity, times the frequency squared.  

Another effect of couplings between modes is to induce \emph{line broadenings and shifts}. This is incompatible with the use of IR line intensities, which are $\delta$ functions at the normal mode frequencies. A way around this obstacle is the recourse to the computation of the time variations of the total \emph{electric dipole moment} of the molecule. The Fourier analysis of these variations deliver the emission spectrum. The Appendix shows that the emission at frequency $\nu$ scales with the product of initial energy content times $\nu^{3}$, times the square of the dipole moment component at frequency $\nu$ (eq. 4). This is proposed here as the more accurate of the three estimates; it comes, of course, at a price: more elaborate and time consuming computational codes. 

For clarity, the three approaches are first illustrated, here, by chemical computations on a small hydrocarbon molecule. This is barely enough to unveil some essential differences between them, as well as a small line broadening. However, our main purpose is to study the formation of \emph{bands and plateaus}, starting from the IR \emph{lines}. This requires larger and larger molecules. Therefore, the dipole moment procedure was then applied to molecules containing, respectively, 101 and 493 atoms of the CHON family. Because of the higher density of IR lines, many broadened lines merge together, giving rise to fewer, but wider, spectral features which are precursors of bands. Moreover, the wings of neighboring lines overlap and form \emph{plateaus}, where line densities are highest. No thermal broadening is involved.

Bands and plateaus can also be obtained by associating a large number of smaller molecules. A very rich diversity of intermediate-size molecules can produce  enough \emph{intercalated broadened lines}, packed within a few characteristic spectral regions, for their sum to look like a continuous band  spectrum. This procedure is illustrated here by a spectrum resulting from the summation of the spectra, computed according to the dipole moment method, of 22 different CHONS structures, ranging from 27 to 112 atoms in size and belonging to 5 families of structures (PAH, aromatics, aliphatic chains, etc.). The time required for the computation of one structure on a desk-top PC is a few hours.

Beyond about 20 $\mu$m, large molecules exhibit far IR bands associated with \emph{bulk} vibration modes (phonons) involving motions of the whole, or large parts, of the molecular structure rather than only the few atoms comprised in the functional groups which carry the mid-IR bands. As the molecule grows larger, it emits more and more in these bands and, relatively, less and less in the UIB bands.

These general computational results could not be obtained on the basis of the usual normal mode analysis, although the latter delivers essential guiding informations in the quest for candidate models of the UIB carriers. 

The so-called \emph{semi-empirical} dynamics code used here takes care of the excitation process and of the anharmonicity of the motion, which depends upon the molecule, the excitation energy and the particular vibration mode considered. If the excitation energy is high enough, overtone and combination bands automatically emerge from the analysis of the variations of the dipole moment in time. This analysis delivers the emission without going through the Normal Modes. No assumptions are added to shift or broaden the bands artificially. Plateaus arise naturally from overlapping band wings. Each of the strong UIBs beyond 5 $\mu$m can be associated with one family of structures.

While DFT (Density Functional Theory) and its derivatives also deliver accurate Normal Mode intensities and frequencies, they are basically \emph{static} methods, so they are not directly applicable to analysis of dipole moment variations. To my knowledge, even Time-Dependent DFT (TDDFT) has been applied only to adiabatic transitions between equilibrium states, not to \emph{dynamics}.

\section{Appendix: Radiated energy}

Present chemistry simulation codes do not include spontaneous radiation. A simple analytical argument can, however, be made to deliver an operational approximation to reality.

When energy is deposited in a molecule, several modes are excited; which ones depends on the excitation process. If the modes were strictly harmonic and, therefore, completely uncoupled, each and everyone would start radiating in the IR, at its own rate, different than the others. Assuming the excitation rate to be slow enough, each mode would ultimately radiate the energy which was initially deposited in it; the radiated power would be obtained by multiplying this with the frequency of excitation events. But, in a real molecule above its ground state, anharmonicity cannot be overlooked and the modes are coupled more or less. As a consequence, intramolecular vibration redistribution then quickly changes the initial distribution into a new one, which is mostly independent of the excitation process, but characteristic of the molecule itself. For simplicity, assume first that the molecule has only 2 vibration modes, and let this characteristic distribution run like $c_{i}^{-1}$, with $i=1,2$, where $c_{i}$ is a pure, constant number. Let 
$E_{i}$ be the energy, $\tau_{i}$ the radiative lifetime of the i\emph{th} mode, and $e$ the energy exchanged between modes. Then,
\begin{eqnarray}                                            
\dot E_{1}= -E_{1}/\tau_{1}+\dot e \\
\dot E_{2}= -E_{2}/\tau_{2}-\dot e \\
\dot e=a(c_{1}E_{1}-c_{2}E_{2}),
\end{eqnarray}
where the dot designates the derivative and $a$ is the mode coupling constant. Note that, when $E_{i}\propto c_{i}^{-1}$, energy exchange stops. It is readily verified that the solution of this set is of the form
\begin{eqnarray} 
E_{1}=\alpha_{1}exp(-t/\tau)+\beta_{1}exp(-t/\sigma) \\
E_{2}=\alpha_{2}exp(-t/\tau)+\beta_{2}exp(-t/\sigma) \\
e=\alpha exp(-t/\tau)+\beta exp(-t/\sigma).
\end{eqnarray}
Identifying separately the coefficients of exponentials, and applying initial conditions, it is found that
\begin{eqnarray} 
c_{1}\alpha_{1}=c_{2}\alpha_{2} \\
\beta_{1}=-\beta_{2},
\end{eqnarray}
so that
\begin{eqnarray} 
\alpha_{1}=c_{2}\frac{E_{01}+E_{02}}{c_{1}+c_{2}} \\
\alpha_{2}=c_{1}\frac{E_{01}+E_{02}}{c_{1}+c_{2}} \\
\beta_{1}==-\beta{2}=E_{01}-c_{2}\frac{E_{01}+E_{02}}{c_{1}+c_{2}} \\
\tau=\frac{\tau_{1}\tau_{2}(c_{1}+c_{2})}{c_{1}\tau_{1}+c_{2}\tau_{2}} \\
\sigma=\frac{1}{a(c_{1}+c_{2})}\,,
\end{eqnarray}
where the subscript $0i$ designates the initial value of the mode energy.

The meaning of these equations is clear when, as is usually the case, the coupling constant, $a$, is strong enough that $\sigma\ll\tau$. In a first, short, phase (the transient, with time constant $\sigma$), radiation is insignificant, but the energies in the two modes are forced to quickly readjust so as to reduce $\dot e$ (eq. 8). They then can relax at the same rate, $\tau$, intermediate between the two radiation rates $\tau_{1}$ and $\tau_{2}$, while energy is being continuously transfered from the less radiative to the more radiative mode, at a very slow rate. 

Finally, the energies radiated by modes 1 and 2 are, respectively
\begin{eqnarray} 
\int E_{1}dt/\tau_{1}=\tau \alpha_{1}/\tau_{1}=\frac{c_{1}+c_{2}}{c_{1}\tau_{1}+c_{2}\tau_{2}}\tau_{2}\alpha_{1} \\
\int E_{2}dt/\tau_{2}=\tau \alpha_{2}/\tau_{2}=\frac{c_{1}+c_{2}}{c_{1}\tau_{1}+c_{2}\tau_{2}}\tau_{1}\alpha_{2}\,,
\end{eqnarray} 
whose sum equals $\alpha_{1}+\alpha_{2}=E_{01}+E_{02}$, as it should, since there is no outlet for the initial energies other than radiation. The ratio of total energies radiated at the respective mode frequencies is 

\begin{equation}
\frac{\alpha_{1}}{\alpha_{2}}\frac{\tau_{2}}{\tau_{1}}\,.
\end{equation}

The forms of the solutions above are easily generalized to any number of modes, assuming a uniform coupling constant. The coupling constant only impacts the transient time constant, $\sigma$, which remains very short anyway (a few tens of picoseconds). One can therefore identify $E_{0i}$ with $\alpha_{0i}$, the characteristic energy distribution; the ratio of radiated energies then become

\begin{equation}
\frac{E_{01}}{E_{02}}\frac{\tau_{2}}{\tau_{1}}\,.
\end{equation}
It is useful to relate the radiative time constants to the IR intensities delivered by the chemistry codes, so as to expedite the computation of the emission spectrum. Indeed, the radiative lifetime is the inverse of Einstein's spontaneous emission coefficient (see Wilson et al. 1955)
\begin{equation}
A(\nu)=\frac{64\pi^{4}\nu^{3}}{3hc^{3}}m(\nu)^{2}\,,
\end{equation}
in electrostatic CGS units, and written in simplified form, for one particular transition and overlooking degeneracies. Here, $m(\nu)$ is the dipole moment oscillating component at frequency $\nu$. Now, it comes as no surprise (see Atkins 1997) that the corresponding IR intensity is also linked to
 $m(\nu)$ by
\begin{equation}
I(\nu)=2.6\nu m(\nu)^{2}\,,
\end{equation}
where $m$ is in Debye and $\nu$ in cm$^{-1}$. Thus, Einstein's coefficient scales like $\nu^{2}I(\nu)$ and $\nu^{3}m(\nu)^{2}$. Finally, the energy ultimately radiated by mode $\nu$ scales like

$E^{*}\propto\nu^{3}m(\nu)^{2}E_{0}(\nu)\,,$

where $E_{0}(\nu)$ is the energy initially deposited in that mode (that is, immediately after the very quick redistribution of the total energy provided by the exciting process). This is the expression used in Sec. 2.3. It is a correction to the presentation of the dipole moment spectrum proposed in Papoular \cite{pap01}.

It should be noted that Einstein's coefficients are generally unknown for the proposed candidate carriers, and their determination is far from trivial. Previous authors in this field have turned around the difficulty by pointing out their linear relation with the corresponding absorption cross-sections. Even those are very scarce, having been obtained by measurements on a few particular molecules. By contrast, the procedure outlined here is easily applicable to any structure, provided it can be handled by the computer at hand.

\end{document}